\documentclass{kapproc} 
\usepackage{psfig}
\kluwerbib

\begin{document}

\articletitle[Spectrophotometry of CNSFRs]
{Optical and Near-IR\\
Spectrophotometry of CNSFR's:\\Preliminary Results\\}

\author{Marcelo Castellanos}
\affil{Grupo de Astrof\'\i sica, Universidad Aut\'onoma de Madrid (Spain)\\}
\email{marcelo.castellanos@uam.es}
\author{Angeles I. D\'\i az}
\affil{Grupo de Astrof\'\i sica, Universidad Aut\'onoma de Madrid (Spain)\\}
\author{Mar Alvarez-Alvarez}
\affil{Grupo de Astrof\'\i sica, Universidad Aut\'onoma de Madrid (Spain)\\}
\author{Elena Terlevich}
\affil{INAOE, Puebla (Mexico)\\}
\chaptitlerunninghead{Spectrophotometry of CNSFRs}

\begin{abstract}
Preliminary results from spectrophotometric observations of Circumnuclear
Star Forming Regions (CNSFRs) in the spiral galaxies NGC 2903, NGC 3351 and
NGC 3504 are presented. The analysis is focused both on the determination
of elemental abundances in the gas and the unveiling of the ionising stellar population responsible of the observed excitation in these regions.
\end{abstract}


\section{Introduction}
The study of CNSFRs in spiral galaxies is rather difficult due to several factors, {\it e.g.} their proximity to the nuclear region (galactocentric distances smaller than 1 Kpc) and their tendency to gather in ring-shaped patterns around the nucleus. Both facts make necessary to deal with moderate-high spatial resolution observations. These circumnuclear HII regions show, in average, high H$\alpha$ luminosities, typical of disk supergiant HII regions (L(H$\alpha$) $\geq$ 10$^{39}$ erg/s)(Kennicutt 1983). These H$\alpha$ luminosities imply the presence of massive ionising stellar clusters ({$>$} 10$^5$ M$_{\odot}$). Therefore, the presence of massive stars, {\it e.g.} Wolf-Rayet (WR) stars, powering these regions should be expected. However, only one spectroscopic detection of WR stars has been reported in CNSFRs (region A in  M83, Bresolin \& Kennicutt 2002). Probably, the success in the detection of WR features in CNSFRs may depend on an adequate signal-to-noise ratio ($\sim$ 30). In order to investigate these caveats, as well as the ionisation structure inside these regions, 12 CNSFRs have been observed in the spiral galaxies NGC 2903, NGC 3351 and NGC 3504. Here we present preliminary results on regions R1 and R2 in NGC 2903.
\section{Observations}
 These spectrophotometric observations were obtained with the 4.2m William Herschel Telescope at the Roque de los Muchachos Observatory, in 2001 January 27, using the ISIS double spectrograph, with the EEV12 and TEK4 detectors in the blue and red arm respectively. The incoming light was split by the dichroic at $\lambda$7500 {\AA}. Two gratings were used: R300B in the blue arm and R600R in the red arm, covering 3400 {\AA} in the blue ($\lambda$3600 to $\lambda$7000) and 800 {\AA} in the near-IR ($\lambda$8850 to $\lambda$9650). This configuration yields spectral resolutions of $\sim$2.0 {\AA} and 1.5 {\AA} FWHM in the blue and red arms respectively, for a slit width of $\sim$ 1 arcsec. The nominal spatial sampling is 0.4 arcsec/pix in both frames and the average seeing for this night was 1.2 arcsec.\\ 
Regions R1 and R2 in NGC 2903 are located at 3 arcsec north and 2 arcsec west from the nucleus. These regions are separated each other 1 arcsec on H$\alpha$ images, hence they could not be resolved, given our higher average seeing. Therefore, throughout the paper, the integrated properties of R1+R2 will be analysed. Blue and near-IR spectra are shown in Figure 1 (upper figures).
\section{Results and Discussion}
The estimation of the reddening constant has been made from the observed intensities of the H$\alpha$, H$\beta$ and H$\gamma$ Balmer lines, which are less unaffected by the presence of an underlying stellar population. H$\delta$ is clearly affected by this stellar absorption. Therefore, observed and theoretical Balmer line intensities (case B and T$_e$ = 5000 K) have been fitted in order to derive the reddening constant. The derived value is c(H$\beta$) = 0.5 which is fully consistent with the derived one from the Paschen recombination lines (0.45). The main reddening corrected line intensities are given in Table 1. Regarding the H$\alpha$ luminosity, the derived value for an adopted distance to this galaxy of 8.6 Mpc, is L(H$\alpha$) = 2.2$\times$10$^{39}$ erg/s, typical of disk supergiant HII regions. This value implies a number of hydrogen ionising photons of Q(H) =1.6$\times$10$^{51}$ s$^{-1}$, (dust-free case and no leakage of ionising radiation), thus requiring ionising cluster masses of 0.5$\times$10$^{6}$ M$_{\odot}$. The derived H$\alpha$ luminosity is in very good agreement with previous CCD photometric observations (P\'erez-Olea 1996). The mean gas density, derived from the [SII]$\lambda$6717, 6731 {\AA} line ratio, is 300 cm$^{-3}$, three times higher than the typical values found for disk giant HII regions.
\begin{figure}[ht]
\begin{minipage}[c]{10mm}
\psfig{figure=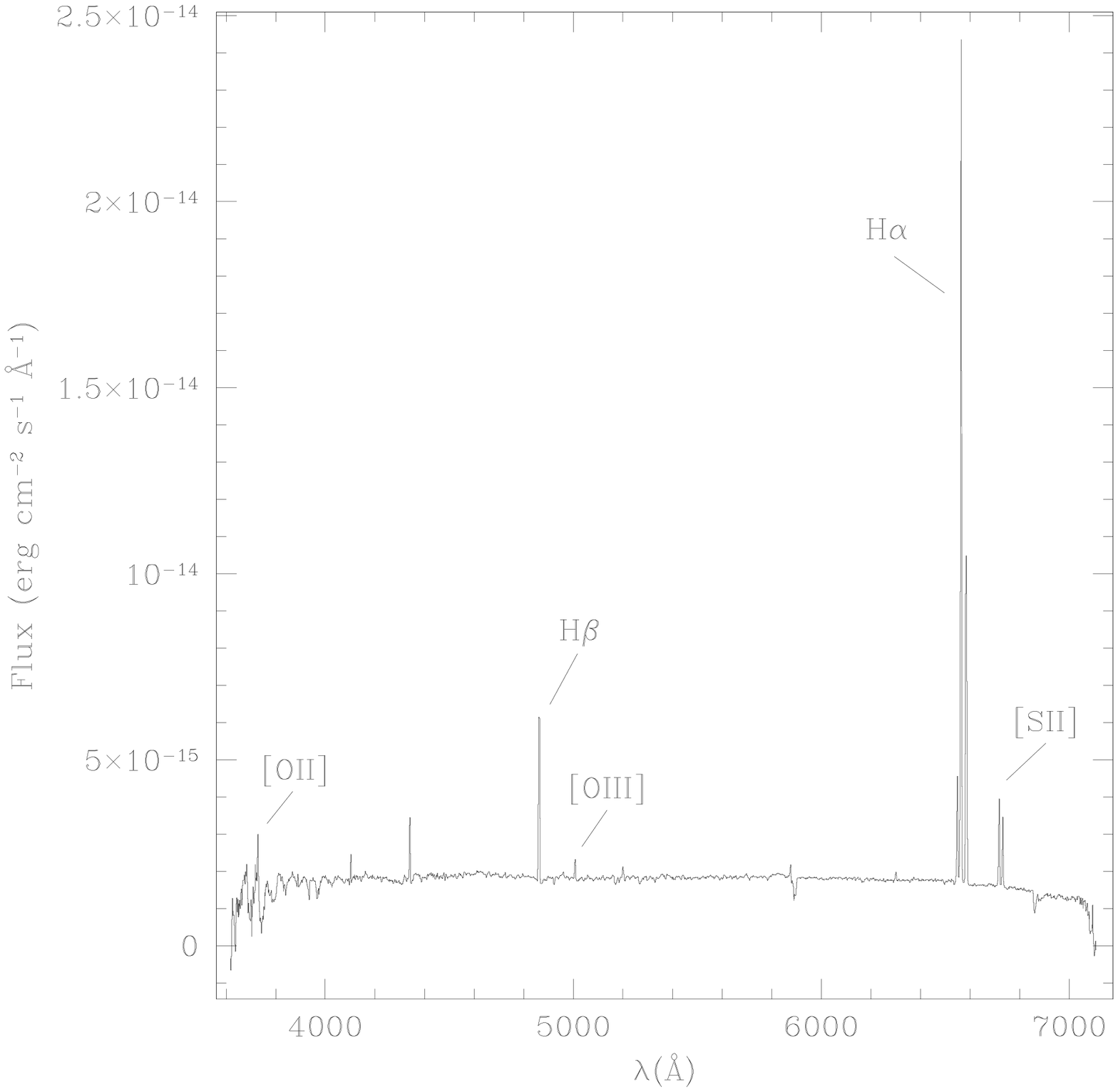,width=6cm,height=6cm,clip=}
\end{minipage}
\begin{minipage}[c]{10mm}
\psfig{figure=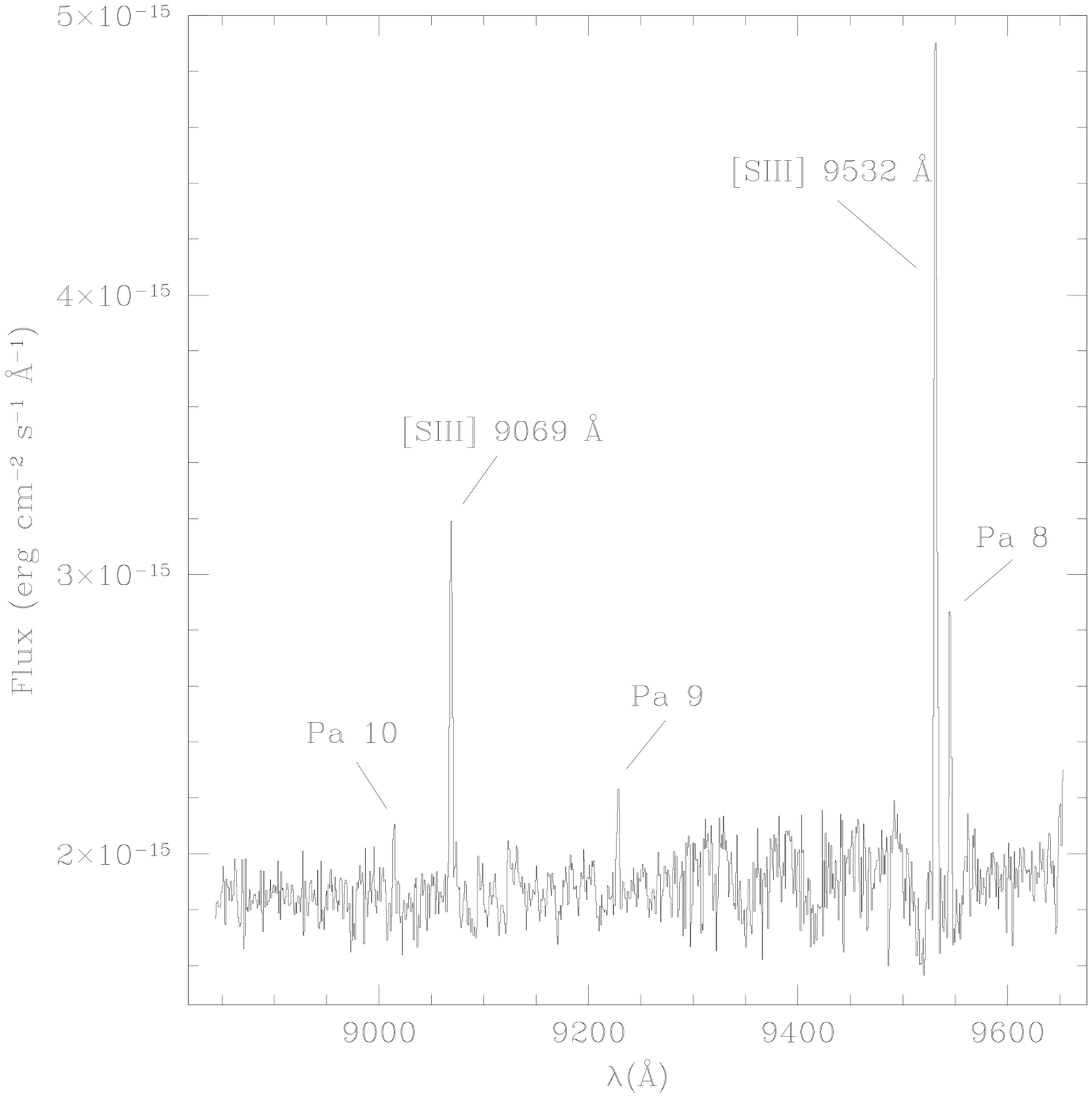,width=6cm,height=6cm,clip=}
\end{minipage}
\begin{minipage}[c]{10mm}
\psfig{figure=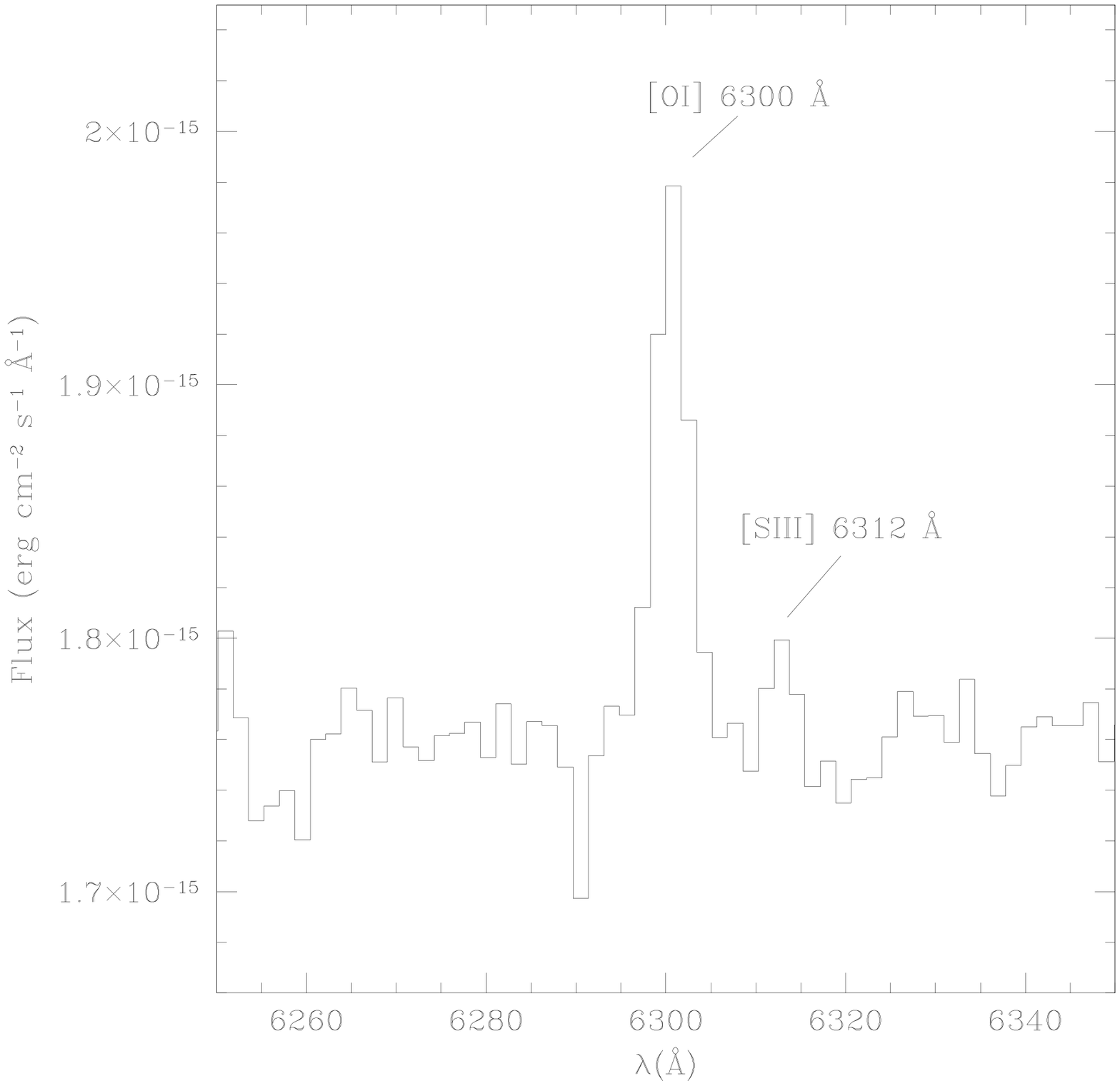,width=6cm,height=6cm,clip=}
\end{minipage}
\begin{minipage}[c]{10mm}
\psfig{figure=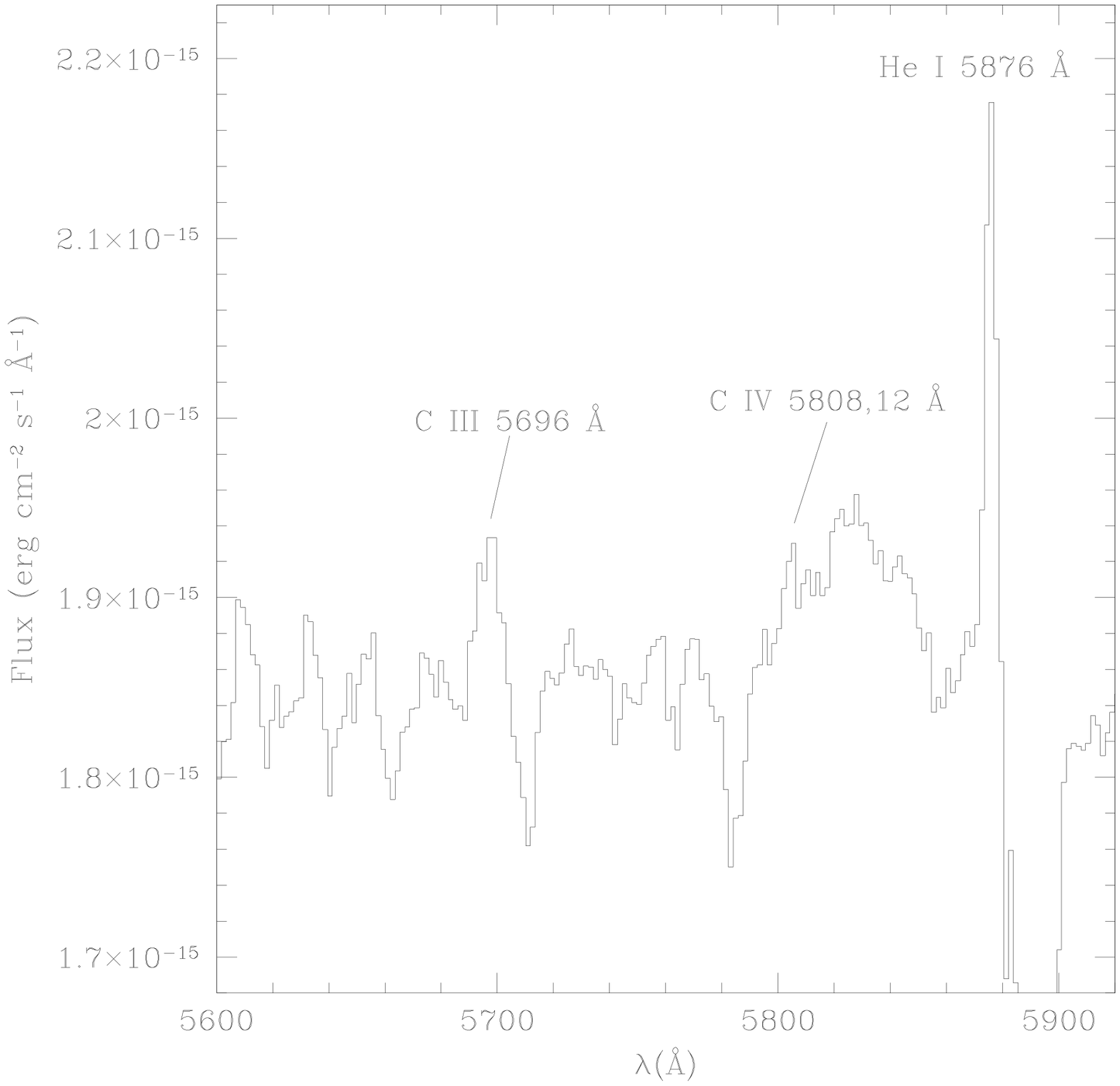,width=6cm,height=6cm,clip=}
\end{minipage}
\caption{Spectral features of region R1+R2 in NGC2903.}
\end{figure}
\begin{table}[ht]
\caption[Reddening corrected line intensities relative to H\$beta$]
{Reddening corrected line intensities relative to H$\beta$ for regions R1+R2}
\begin{tabular*}{\textwidth}{@{\extracolsep{\fill}}lllllll}
\sphline
\it Line&[OII]&[OIII]&[NII]&[SII]&[SII]&[SIII] \cr
\it $\lambda$(\AA)&3727&5007&6584&6717&6731&9069 \cr
\sphline
I($\lambda$)/I(H$\beta$)&0.40&0.10&1.24&0.31&0.26&0.08 \cr
\sphline
\end{tabular*}
\end{table}
According to D\'\i az \& P\'erez-Montero (2001), those regions with values for log O$_{23}$ $\leq$ 0.47 and -0.5 $\leq$ log S$_{23}$ $\leq$ 0.28 must have solar (12 + log(O/H) = 8.92) or oversolar metal content. Region R1+R2 satisfies these conditions (log O$_{23}$ = -0.27 and log S$_{23}$ = -0.07), hence it can be considered a high metallicity HII region on the basis of empirical calibrations. On the other hand, the weak [SIII]$\lambda$6312 {\AA} auroral forbidden line has been clearly detected (see Figure 1, bottom left), with a reddening corrected intensity of 0.004 relative to H$\beta$. This value, together with the near-IR forbidden sulphur line intensities, yields an electron temperature of $\sim$ 8500 K. This value is around 3000 K higher than expected for a region of high metal content. \\
Another interesting feature in our observed spectrum is the presence of a broad 5696 {\AA} C III emission (see Figure 1, bottom right), which constitutes a clear signature for the presence of WC type Wolf-Rayet stars. The presence of these stars is expected in high metallicity environments (Conti \& Vacca 1990), which is the case for our observed region. The detection of 5808 {\AA} C IV emission is also reported. The blue Wolf-Rayet bump ($\sim$ 4660 {\AA}) is also present to a lesser extent. The number of WC stars responsible of the observed luminosity at 5696 {\AA} and 5808 {\AA} can be derived by assuming that WN stars do not contribute to these lines. At any rate, this calculation is rather uncertain because it is difficult to establish the dominant WC subtype. This fact affects the adopted average luminosity in these lines (see Pindao et al. 2002, for a review of this matter).\\
Absorption features in the integrated spectrum are also observed (K \& H Ca II $\lambda\lambda$3933,3967 {\AA},  G band at $\lambda$4301 {\AA}, Mg I $\lambda$5175 {\AA}). Old (Gyr) populations of the galactic bulge could be responsible for these lines.

\end{document}